\documentclass[12pt,preprint]{aastex}

\begin{document}

\def\wisk#1{\ifmmode{#1}\else{$#1$}\fi}

\def\lt     {\wisk{<}}
\def\gt     {\wisk{>}}
\def\le     {\wisk{_<\atop^=}}
\def\ge     {\wisk{_>\atop^=}}
\def\lsim   {\wisk{_<\atop^{\sim}}}
\def\gsim   {\wisk{_>\atop^{\sim}}}
\def\kms    {\wisk{{\rm ~km~s^{-1}}}}
\def\Lsun   {\wisk{{\rm L_\odot}}}
\def\Zsun   {\wisk{{\rm Z_\odot}}}
\def\Msun   {\wisk{{\rm M_\odot}}}
\def\um     {$\mu$m}
\def\mic     {\mu{\rm m}}
\def\sig    {\wisk{\sigma}}
\def\etal   {{\sl et~al.\ }}
\def\eg     {{\it e.g.\ }}
 \def\ie     {{\it i.e.\ }}
\def\bsl    {\wisk{\backslash}}
\def\by     {\wisk{\times}}
\def\half {\wisk{\frac{1}{2}}}
\def\third {\wisk{\frac{1}{3}}}
\def\nwm2sr {\wisk{\rm nW/m^2/sr\ }}
\def\nw2m4sr {\wisk{\rm nW^2/m^4/sr\ }}

\title{A new measurement of the bulk flow of X-ray luminous clusters of galaxies}

\author{A. Kashlinsky\altaffilmark{1}, F.  Atrio-Barandela\altaffilmark{2},
 H.  Ebeling\altaffilmark{3}, A. Edge\altaffilmark{4}, D. Kocevski\altaffilmark{5}}
\altaffiltext{1}{SSAI and Observational Cosmology Laboratory, Code
665, Goddard Space Flight Center, Greenbelt MD 20771;
alexander.kashlinsky@nasa.gov} \altaffiltext{2}{Fisica Teorica,
  University of Salamanca, 37008 Salamanca, Spain}
   \altaffiltext{3}{Institute for Astronomy, University of Hawaii, 2680
  Woodlawn Drive, Honolulu, HI 96822} \altaffiltext{4}{Department of Physics,
University of Durham, South Road, Durham DH1 3LE, UK }
\altaffiltext{5}{Department
  of Physics, University of California at Davis, 1 Shields Avenue, Davis, CA
  95616}


\begin{abstract}
{We present new measurements of the large-scale bulk flows of
galaxy clusters based on 5-year WMAP data and a significantly
expanded X-ray cluster catalogue. Our method probes the flow via
measurements of the kinematic Sunyaev-Zeldovich (SZ) effect
produced by the hot gas in moving clusters. It computes the dipole
in the cosmic microwave background (CMB) data at cluster pixels,
which preserves the SZ component while integrating down other
contributions. Our improved catalog of over 1,000 clusters enables
us to further investigate possible systematic effects and, thanks
to a higher median cluster redshift, allows us to measure the bulk
flow to larger scales.  We present a corrected error treatment and
demonstrate that the more X-ray luminous clusters, while fewer in
number, have much larger optical depth, resulting in a higher
dipole and thus a more accurate flow measurement. This results in
the observed correlation of the dipole derived at the aperture of
zero monopole with the monopole measured over the cluster central
regions. This correlation is expected if the dipole is produced by
the SZ effect and cannot be caused by unidentified systematics (or
primary cosmic microwave background anisotropies). We measure that
the flow is consistent with approximately constant velocity out to
at least $\simeq$800 Mpc. The significance of the measured signal
peaks around 500 $h_{\rm 70}^{-1}$Mpc, most likely because the
contribution from more distant clusters becomes progressively more
diluted by the WMAP beam. We can, however, at present not rule out
either that these more distant clusters simply contribute less to
the overall motion.}

\end{abstract}
\keywords{Cosmology - cosmic microwave background - observations -
diffuse radiation - early Universe}


The large-scale isotropy of the Universe and the small-scale
inhomogeneities that evolved into galaxies are thought to
originate during inflationary expansion in the early Universe.
Inflation posits that the primeval space-time was inhomogeneous
and this structure should have been preserved on sufficiently
large scales. At minimum large-scale peculiar velocities should
arise from gravitational instability caused by mass
inhomogeneities seeded during the inflationary expansion. On
scales $\gsim$100Mpc the standard inflationary scenario leads to
robust predictions for these velocities. There the initial
Harrison-Zeldovich slope of the mass fluctuations is preserved, so
peculiar velocities induced by gravitational instability must
decrease linearly with scale (e.g. Kashlinsky \& Jones 1991); for
the concordance $\Lambda$CDM model $V_{\rm rms} \sim
250(\frac{100h^{-1}{\rm Mpc}}{d})$ km/sec at $d{>}$50--100$h^{-1}$
Mpc.

Our discovery of a coherent large-scale flow of galaxy clusters
with significantly larger amplitude than expected out to $\simeq
300$Mpc (Kashlinsky et al 2008, 2009 -- KABKE1,2) represents a
challenge to the gravitational instability paradigm. Such a "dark
flow" could indicate a tilt created by the pre-inflationary
inhomogeneous structure of space-time (Turner 1991, Grishchuk
1992, Kashlinsky et al 1994, KABKE1) and might provide an indirect
probe of the Multiverse. Various explanations have been put
forward, including that the flow points to a higher-dimensional
structure of gravity (Afshordi et al 2009, Khoury \& Wyman 2009),
or that it reflects the pre-inflationary landscape produced by
certain variants of string cosmology (Mersini-Houghton \& Holman
2009, Carrol et al 2008).

Making use of an expanded cluster catalog and deeper WMAP
observations, we have worked to verify, and expand the Dark Flow
study through a program we have dubbed {\it SCOUT} ({\bf
S}unyaev-Zel'dovich {\bf C}luster {\bf O}bservations as probes of
the {\bf U}niverse's {\bf T}ilt). First results from this
experiment are reported here.

\section{Data and analysis}

KABKE1,2 and this work utilize a method of Kashlinsky \&
Atrio-Barandela (2000, hereafter KA-B), which measures CMB dipole
at the locations of X-ray clusters. When averaged over many
isotropically distributed clusters moving at a significant bulk
flow with respect to the CMB, the kinematic term dominates the SZ
signal, thereby enabling a measurement of $V_{\rm bulk}$ over that
distance. In Atrio-Barandela et al (2008, hereafter AKKE) we
demonstrated that 1) the thermal SZ (TSZ) signal from clusters
extends well beyond the measured X-ray extent $\Theta_{\rm X}$,
and 2) the intra-cluster gas distribution is well approximated by
the NFW profile (Navarro et al 1996) expected for dark matter in a
$\Lambda$CDM model. The temperature $T_X$ of hot gas distributed
according to these profiles decreases significantly from the
cluster cores to the cluster outskirts (Komatsu \& Seljak 2001),
consistent with current measurements (Pratt et al 2007) and
numerical simulations (e.g. Borgani et al 2004). Consequently, the
monopole produced by the TSZ component ($\propto \tau T_X$)
decreases, as we increase the cluster aperture, whereas the dipole
due to the kinematic SZ (KSZ) component ($\propto \tau$, the
optical depth due to Thomson scattering) remains measurable out to
the aperture where we still detect the TSZ decrement in unfiltered
maps (KABKE2). As in KABKE1,2 our dipole coefficients are
normalized such that the dipole power $C_{1}$ due to a coherent
motion at velocity $V_{\rm bulk}$ is $C_{1,{\rm kin}}= T_{\rm
CMB}^2 \langle \tau \rangle^2 V_{\rm bulk}^2/c^2$, where $T_{\rm
CMB} =2.725$K. We adopt $\Omega_{\rm total}=1, \Omega_\Lambda=0.7,
H_0=70h_{70}$ km/sec/Mpc.

To improve upon the all-sky cluster catalogue of Kocevski \&
Ebeling (2006) used by KABKE1,2, we have screened the ROSAT
Bright-Source Catalogue (Voges et al.\ 1999) using the same X-ray
selection criteria (including a nominal flux limit of $1\times 10^
{-12}$ erg/sec, 0.1--2.4 keV) as employed during the Massive
Cluster Survey (MACS, Ebeling et al.\ 2001), as well as the same
optical follow-up strategy. Unlike MACS, we apply neither a
declination nor a redshift limit though, thereby creating an
all-sky list of cluster candidates that extends to three times
fainter X-ray fluxes than in KABKE1,2. Optical follow-up
observations of clusters identified in this manner and lacking
spectroscopic redshifts in the literature (and MACS) are well
underway using telescopes on Mauna-Kea/Hawai`i and La-Silla/Chile.
As a result, our interim all-sky cluster catalogue currently
comprises in excess of 1,400 X-ray selected clusters, {\it all} of
them with spectroscopic redshifts. X-ray properties of all
clusters (most importantly total luminosities and central electron
densities) are computed as before (KABKE2). Within the same
$z$-range ($z\la0.25$) as our previous study, our new catalog
comprises 1,174 clusters outside the KP0 CMB mask. To eliminate
low-mass galaxy groups we require that clusters feature $L_X\geq 2
\times 10^{43}$ erg/sec; 985 systems meet this criterion. This
sample represents a significant improvement over the one in
KABKE1,2 largely because of the substantial increase ($z_{\rm
median}<0.1$ in KABKE1,2 vs $z_{\rm median}\simeq 0.2$ below) in
median cluster redshift and the higher fraction of intrinsically
very X-ray luminous systems (Fig. 1a).

We applied this new cluster sample to the 5-year WMAP CMB data,
processed as described in detail in KABKE1,2. The all-sky dipole
in the foreground-cleaned maps from
http://www.lambda.gsfc.nasa.gov was removed, and standard CMB
masking was applied.
We also need to remove the primary CMB fluctuations, produced at
last scattering, as they are highly correlated and would
contribute significantly to the measured dipole. To this end, all
maps were filtered as in KABKE1,2 with a filter that minimizes
$\langle (\delta T - \delta T_{\Lambda{\rm CDM}})^2\rangle$. The
error budget associated with our filtering is discussed by
Atrio-Barandela et al (2010, hereafter AKEKE).

Measurement errors were computed as in KABKE1,2 with one important
correction. Although the filtering removes much of the CMB
fluctuations, a residual component remains, due to cosmic variance
and imperfections of the theoretical model. Since this residual is
common to all WMAP bands, a component of the errors is correlated
between the various DA maps. We address this issue in the
following manner. For each of the eight differential assemblies
(DAs) we simulate 4,000 realizations with $N_{\rm cl}=$100--1,000
randomly selected pseudo-clusters outside of the cluster pixels
and the CMB mask. In each realization we select the {\it same}
pseudo-clusters for all DAs and evaluate the {\it mean} monopole
and dipole averaged over all DAs: $\bar{a}_0, \bar{a}_{1m}$. For
the 4,000 realizations at each $N_{\rm cl}$ we compute the mean
and dispersion of $\bar{a}_0, \bar{a}_{1m}$ over all the
realizations. The distributions have zero mean and their
dispersion gives errors which scale as $N_{\rm cl}^{-1/2}$. We
find to good accuracy that the distribution of the simulated
dipoles (and monopoles) is Gaussian and the errors on each of the
averaged dipole components are $\sigma_{1m} \simeq 15
\sqrt{3/N_{\rm cl}}\mu$K and on the monopole $\sigma_0 \simeq 15
\sqrt{1/N_{\rm cl}} \mu$K as explained in great detail in AKEKE;
the errors of the $x/z$ dipole component are slightly
larger/smaller because of the Galactic mask (Fig. 1b).

\section{Results}

We used the filtered maps to  compute the dipole and monopole
terms, $a_{1m}^i,a_0^i$ for each DA of the WMAP Q, V, W bands
($i=1,...,8$) for clusters in cumulative redshift bins up to a
given $z$. The results were averaged to obtain the mean values
over all eight DAs, $\bar{a}_{1m},\bar{a}_0$. Since the volume
probed out to low $z$ is too small for meaningful measurements,
Table 1 lists results only for $z$-bins with sufficient
signal-to-noise ratios (S/N). In KABKE1,2 we computed the dipole
in progressively increasing apertures but no larger than
30$^\prime$ to prevent geometric biases from very nearby Coma-type
clusters. To further reduce any selection effect related to the
apparent X-ray extent measured by ROSAT, we here impose a {\it
constant} aperture for {\it all clusters} (see Table 1) and
compute the dipole component for each $z$-bin at the constant
aperture at which the monopole (initially negative because of the
TSZ component) vanishes.

The improved cluster catalog allows us to extend our study to
higher $z$,  and to further test the impact of systematics. We
create $L_{\rm X}$-limited subsamples which achieves two important
objectives. 1) As the $L_X$ threshold is raised, fewer clusters
remain and the statistical uncertainty of the dipole increases
($\propto 1/\sqrt{N_{\rm cl}(L>L_X)}$). If, however, all clusters
are part of a bulk flow of a given velocity $V_{\rm bulk}$, very
X-ray luminous clusters will produce a larger CMB dipole ($\propto
\tau V_{\rm bulk}$), an effect that might overcome the reduced
number statistics, giving a higher S/N in the measured dipole. 2)
Since, as outlined under (1), the dipole signal should increase
with cluster luminosity, whereas systematic effects can be
expected to be independent of $L_{\rm X}$, an actual observation
of such a correlation would lend strong support to the validity of
our measurement and the reality of the "dark flow".

Fig. 1a shows that the depth to which we probe the flow increases
dramatically as the $L_X$-threshold is raised. Table 1 shows the
results for each subsample. The monopole is strongly negative in
the smallest apertures due to the dominance of the TSZ component
in the central regions and the dipole is shown at the aperture
where monopole vanishes. Of the three dipole components, the
$y$-component is best determined, its value always remaining
negative and its S/N increasing strongly with increasing $L_X$. As
shown in Fig. 1c the amplitudes of the latter and of the monopole
in the central parts are strongly, and approximately linearly,
correlated. {\it This correlation provides strong evidence against
unknown systematics systematics - or primary CMB - causing our
measurement}.

Table 1 quantifies our finding of a statistically significant
dipole out to the largest scales probed ($\sim
800h_{70}^{-1}$Mpc). In KABKE2 we discussed in detail why this
dipole is unlikely to be produced by systematics; we briefly
revisit the issue here: 1) At high statistical significance the
dipole originates only at cluster positions, and must thus
originate from CMB photons that passed through the hot
intra-cluster gas. For the same reason the dipole can not be due
to a residual contribution from the all-sky CMB dipole. HEALPix
ANAFAST routines (Gorski et al 2005), employed in the analysis,
further remove any all-sky dipole before the filtered maps are
produced. 2) Since the dipole is measured at zero monopole, the
contributions from TSZ  and other cluster emissions to the dipole
are negligible. 3) The variations in the final aperture where the
dipole is measured were very small in KABKE1,2, and the dipole
signal remains in this work which uses a fixed aperture for {\it
all} clusters. So the dipole is not affected by the variations in
cluster $\Theta_X$, which in any case is much smaller than the
final apertures in KABKE1,2. 4)  The measured CMB quadrupole is
significantly different from that of the $\Lambda$CDM model, so a
significant part of the CMB quadrupole is not removed by our
filter and could leak into other multipoles via the mask. However,
we set the filter to zero at $\ell\leq 4$ (KABKE2, AKEKE) for the
final maps. More importantly, we have modified the pipeline to
remove the all-sky quadrupole from the {\it original} maps and
find no noticeable difference in the dipole computed at the
cluster locations. This also removes the fully relativistic
components from the local motion $v_{\rm local}$, down to $(v_{\rm
local}/c)^3$ corrections to the octupole. (5) Finally, we
demonstrate that more luminous clusters make a larger, and
statistically more significant, contribution to the dipole, as is
expected if all clusters participate in the same flow, independent
of $L_X$. Note also that the intra-cluster medium motions from
cluster mergers have random directions and thus average down in
large catalogs, contributing negligibly to the noise budget below.

Although the final dipole is measured at zero monopole, tests of
cross-talk were conducted as in KABKE2 by constructing CMB maps
from TSZ and KSZ components of varying $V_{\rm bulk}$ using the
derived catalog parameters, but randomly placed clusters. The
parameters from randomly placed clusters were compared with those
from the original clusters - the cross-talk effects are small with
results similar to Fig. 6 of KABKE2 (AKEKE). To test the
robustness further we select 4,000 random subsets of clusters
within a given configuration and compute the mean dipole and its
dispersion. The distribution of the dipoles is Gaussian, the
dispersion scales as $N_{\rm cl}^{-1/2}$ and the results are
consistent with all subsets of clusters moving in the same way
within the estimated errors. E.g. randomly selecting 250/150
$L_X\geq2\times10^{44}$erg/sec clusters out of 322/208 at $z\leq
0.25/0.2$ gives $(\bar{a}_{1x},\bar{a}_{1y},\bar{a}_{1z})=
(3.7\pm1.8,-4.1\pm1.7,4.2\pm1.5)/(3.6\pm2.1,-5.7\pm2.3,4.5\pm1.9)\mu$K.

To calibrate our dipole measurement in terms of an equivalent bulk
velocity, we proceed as in KABKE2. This still suffers from a
systematic bias which {\it overestimates} the amplitude of the
velocity. More importantly, we measure the dipole from the
filtered maps, and the convolution of the intrinsic KSZ signal
with the filter can change the sign of the former for NFW
clusters. The TSZ signal, being more concentrated as shown in
Fig.~9 of KABKE2, is less susceptible to this effect. We therefore
currently constrain only the axis of the motion; the direction
along this axis should result from future applications of the KA-B
method particularly to the 217 GHz {\it Planck} data, where the
TSZ component vanishes and the angular resolution is 5$^\prime$, a
good match to the inner parts of clusters at z$\sim$0.1-0.2. Our
present pipeline computes the cluster properties (central electron
density, $n_{e,0}$, $T_X$, core radius $R_{\rm core}$) assuming a
$\beta$-model ($\beta$=2/3). This model has been shown by us to be
deficient at the cluster outskirts and must be replaced by the NFW
profile (AKKE). We hope to accomplish this difficult task in the
future with better CMB (Planck) and X-ray (Chandra/XMM) data. To
summarize, our current calibration may {\it overestimate} the
amplitude of the flow and, strictly speaking, we currently measure
only the axis of motion. We stress, however, that the existence of
the flow itself is not affected by this systematic uncertainty. We
generated CMB temperatures from the KSZ effect for each cluster
and estimate the dipole, $C_{1,100}$, contributed by each 100
km/sec of bulk-flow in each $L_X,z$-bin. Since the $\beta$-model
still gives a fair approximation to cluster properties around
$\Theta_X$, we present in Table 1 the final calibration
coefficients evaluated at apertures of $5^\prime$ and $\Theta_X$
in radius. When averaged over clusters of all X-ray luminosities
the mean calibration is $\sqrt{\langle C_{1,100}\rangle} \simeq
0.3 \mu$K in each of the $z$-bins. Within the uncertainties, the
dependence of the calibration on $L_X$ is in good agreement with
the measured dipoles, particularly for the most accurately
measured $y$-component. Table 1 also shows the mean central
optical depth, $\langle\tau_0\rangle$, evaluated from the cluster
catalog. Its variation with $L_X$ is also in good agreement with
that of the measured dipole, which indicates that the clusters can
indeed be assumed to have similar profiles.

With the calibration factors in Table 1, our results are
compatible with a consistently coherent flow at all $z$. Since the
different cluster subsamples probe different depths ($z_{\rm
mean/median}$), we proceed as follows to isolate the overall flow
across all available scales: for the flow which extends from the
smallest to the largest $z$ in each $z$-bin, we model the dipoles
as $a_{1m}^n = \alpha_n V_m$, where $\alpha_n$ is the calibration
constant for clusters in $n$-th luminosity bin. (Note that for
each $z$-bin the luminosity bins are statistically independent).
Both $\sqrt{C_{1,100}}$ and $\langle \tau_0\rangle$ scale
approximately linearly with the better measured dipole
coefficient, $a_{1,y}$, as they should in case of a coherent
motion; the linear correlation coefficients are $r=$0.92/0.93 for
correlation of $-a_{1,y}$ with columns 7/8. We then compute by
regression the velocity components with their uncertainties using
the $L_X$-divisions at each $z$-bin. These numbers are shown in
the summary row following each $z$-bin quantities for
$\alpha=\sqrt{C_{1,100}}$ at $5^\prime$. We also computed them for
$\alpha_n$ given by $\sqrt{C_{1,100}}$ at $\Theta_X$ and by
$\langle\tau_0\rangle$ in column (7) normalized to the observed
mean value of $\langle C_{1,100}\rangle$. Both give results
essentially identical to the ones shown in the Table. The latter
approximation is equivalent to assuming that all clusters have
universal profiles, so that the final effective optical depth is
$\propto\langle\tau_0\rangle\times$(reduction factor). The results
here are consistent with KABKE1,2 measurements on smaller scales
($\lsim400$Mpc), which with {\it revised} errors become
$(\bar{a}_{1x}, \bar{a}_{1y},\bar{a}_{1z})=(0.7\pm1.2, -3.3 \pm
1.1, 0.5\pm 1.)/(0.6\pm1.2, -2.7 \pm 1.1, 0.6\pm 1.)$ for $z\leq
0.2/0.3$ with $\sqrt{C_{1,100}}\simeq 0.3\mu$K (Table 2, KABKE2).

The dipoles are larger for greater $L_X$ clusters consistent with
the dipole originating from the bulk motion of the clusters. We
note the apparent trend in the central values of the better
determined $y$-component peaking at $z\leq 0.16$ and decreasing
towards higher $z$. It is likely that this decrease is due to
dilution of progressively more distant clusters, as shown by their
smaller monopoles. Nevertheless, it is also possible, in
principle, that the flow is dominated by the $z\leq 0.16$ clusters
with the more distant clusters contributing little to the dipole.

\section{Discussion}

We find a high likelihood of the existence of a coherent bulk flow
extending to at least $z\simeq 0.2$ with an amplitude and in a
direction which are in good agreement with our earlier
measurements. Our result constitutes a significant improvement in
that it extends our previous work to approximately twice the
distance accessible to KABKE1,2, supporting their hypothesis that
the flow likely extends across much (or all) of the Hubble volume.
The flow's axis is also consistent with earlier measurements of
the local cluster dipole (Kocevski et al.\ 2004) as well as with
independent measurements of bulk flows on smaller scales by
Watkins et al (2009). The velocity reported there is smaller than
the numbers in Table 1, although the two amplitudes agree at $<
2$-$\sigma$ level. Agreement between the two sets of central
values would require $\sqrt{\langle C_{1,100}\rangle}
\sim$0.4-0.5$\mu$K, or a reduction by a factor of $\sim2$ from
unfiltered values for NFW cluster profiles. Feldman et al (2009)
extend the Watkins et al analysis and find that the absence of
shear in their flow at $\la$50-100Mpc is consistent with the
KABKE1 suggestion of the attractor at superhorizon distances.

Fig. \ref{fig:f2} displays the results obtained in this study
compared to expectations from the concordance $\Lambda$CDM model
for 95\% of cosmic observers.
These results
cast doubt on the notion that gravitational instability from the
observed mass distribution is the sole -- or even dominant  --
cause of the detected motion.
If the current picture is confirmed, it will have profound
implications for our understanding of the global structure of
space-time and our Universe's place in it.

We acknowledge NASA NNG04G089G/09-ADP09-0050 and
FIS2006-05319/GR-234 grants from Spanish Ministerio de Educaci\'on
y Ciencia/Junta de Castilla y Le\'on.

{\bf REFERENCES}\\
Afshordi, N. Geshnizjahi, G. \& Khoury, J. 2009, JCAP, 8, 30\\
Atrio-Barandela, F., Kashlinsky, A., Kocevski, D. \& Ebeling, H.
2008, Ap.J. (Letters), 675, L57 (AKKE)\\
Atrio-Barandela, F., Kashlinsky, A., Ebeling, H., Kocevski, D. \&
Edge, A. 2010, ApJ, submitted, arXiv:1001.1261 (AKEKE)\\
Borgani, S. et al  2004, Mon. Not. R. Astron. Soc., 348, 1078\\
 Carrol, S. et al 2008, arxiv:0811.1086\\
Ebeling, H.; Edge, A. C.; Henry, J. P., 2001, ApJ, 553, 668\\
Feldman, H., Watkins, R. \& Hudson, M.J. 2009, arXiv:0911.5516v1\\
Gorski, K. et al 2005, Astrophys. J., 622, 759\\
 Grishchuk, L. P. 1992, Phys. Rev. D 45,
 4717\\
Kashlinsky, A., Tkachev, I., Frieman, J.  1994, Phys. Rev. Lett.,
73, 1582\\
Kashlinsky, A. \& Atrio-Barandela, F.  2000, Astrophys. J., 536,
L67 (KA-B)\\
Kashlinsky, A., Atrio-Barandela, F., Kocevski, D. \& Ebeling, H.
2008, Ap.J., 686, L49 (KABKE1)\\
Kashlinsky, A., Atrio-Barandela, F., Kocevski, D. \& Ebeling, H.
2009, Ap.J., 691, 1479 (KABKE2)\\
Kashlinsky, A. \& Jones, B.J.T. 1991, Nature, 349, 753\\
Khoury, J. \& Wyman, M. 2009, PhRevD, 80, 064023\\
Kocevski, D.D., Mullis, C.R., \& Ebeling, H. 2004, Astrophys. J.,
608, 721\\
Kocevski, D.D. \& Ebeling, H.  2006, Astrophys. J., 645, 1043\\
Komatsu, E. \& Seljak, U. 2001, Mon. Not. R. Astron. Soc., 327,
1353\\
Mersini-Houghton, L. \& Holman, R. 2009, JCAP,  2, 6\\
Navarro, J.F., Frenk, C.S. \& White, S.D.M. 1996, Astrophys. J.,
462, 563\\
Pratt, G.  et al 2007, Astron. Astrophys. 461, 71\\
Turner, M. S. 1991, Phys.Rev., 44, 3737\\
Watkins, R., Feldman, H. A. \& Hudson, M. J. 2009, MNRAS, 392,
743\\
 Voges, W. et al 1999, A\&A, 349,
389\\

\clearpage
 \thispagestyle{empty}
\begin{deluxetable}{l c c c c c | c | c c | c c c c c}
\tablewidth{0pt} \tabletypesize{\scriptsize} \rotate
\tablecaption{RESULTS  \label{table} } \tablehead{ \colhead{(1)} &
\colhead{(2)} & \colhead{(3)} & \colhead{(4)} & \colhead{(5)} &
\colhead{(6)} & \colhead{(7)} & \colhead{(8)} & \colhead{ } &
\colhead{ } & \colhead{ }
 & \colhead{(9)}  & \colhead{ }  & \colhead{ }  \\
} \startdata
 $z\leq$ & $L_X$-bin & $N_{\rm cl}$ &
$z_{\rm mean}/z_{\rm median}$  & $\bar{a}_{1,x}, \bar{a}_{1,y},
\bar{a}_{1,z}$ & $\sqrt{C_1}$ & $\langle\tau_0\rangle$ &
$\sqrt{C_{1,100}}$ &
($\mu$K) & & & $\bar{a}_0$ & ($\mu$K) & \\
& $10^{44}$ erg/s &  &  & $\mu$K & $\mu$K & $\times10^{-3}$ & 5$^\prime$ & $\Theta_{\rm X}$ & $10^\prime$ & $15^\prime$ & $20^\prime$ & $25^\prime$ & $30^\prime$\\
 \hline
0.12$^{*}$ & 0.2--0.5 & 142 & 0.061/0.060 & $-4.2\pm2.7, -0.7\pm2.3, 0.5\pm2.3$  & $4.3\pm2.7$ & 2.8 & 0.2301 &  0.1942 &  --2.8 & 0.1  & -- & -- & -- \\
0.12 & 0.5--1 & 194 & 0.081/0.082 & $-2.7\pm2.3, -2.3\pm2.0, 1.4\pm2.0$  & $3.9\pm2.2$ & 3.5 & 0.2989 & 0.2561 & --2.4 & --1.2  & --0.1 & 0.6 & 0.8 \\
0.12 & $>1$ & 180 & 0.083/0.086 & $4.9 \pm 2.4, -4.5 \pm 2.1, 1.5\pm 2.0$ & $6.8\pm2.2$ & 5.4 & 0.4610 & 0.3496  & --11.1 & --6.5  & --3.1 & --0.8 & 0.5 \\
\hline \\
\multicolumn{14}{l}{\vspace{0.1in} $d\sim 250-370h_{70}^{-1}$Mpc;
$(V_x,V_y,V_z)= (174 \pm
  407, -849 \pm 351 , 348 \pm 342 )\times\frac{+0.3\mu
    K}{\sqrt{\langle{C}_{1,100}\rangle}}$ km/sec; $V_{\rm Bulk}=(934 \pm 352)\times\frac{0.3\mu K}{\sqrt{\langle{C}_{1,100}\rangle}}$ km/sec; $(l_0,b_0)= (282\pm 34, 22 \pm 20)^\circ$} \\
\hline
 0.16 & 0.5--1 & 226 & 0.089/0.087 & $-1.5\pm2.2, -0.6 \pm1.9, 2.1 \pm 1.8$  & $2.7\pm1.9$ & 3.5 & 0.2843 & 0.2363 & --2.8 & --1.8  & --0.6 & 0.2 & 1.6 \\
0.16 & 1--2 & 191 & 0.106/0.107 & $1.9\pm2.3, -2.8 \pm 2.0, -0.5 \pm 2.0$  & $4.1\pm2.2$ & 4.4 & 0.3480 & 0.2894 & --4.9 & --1.4  & 0.4 & 1.3 & 1.8 \\
0.16 & $>2$ & 130 & 0.115/0.125 & $4.2\pm2.8, -8.0\pm2.4, 4.9\pm2.4$  & $10.3\pm2.5$ & 6.8 & 0.4930 & 0.4238 & --11.7 & --7.1  & --2.9 & --0.3 & 0.8 \\
\hline \\
\multicolumn{14}{l}{$^{(a)}$ $d\sim 370-540 h_{70}^{-1}$Mpc; $(V_x,V_y,V_z) = (410 \pm 379 , -1,012 \pm 326, 566 \pm 319 )\times\frac{+0.3\mu K}{\sqrt{\langle{C}_{1,100}\rangle}}$ km/sec; $V_{\rm Bulk}=(1,230\pm 331 )\times\frac{0.3\mu K}{\sqrt{\langle{C}_{1,100}\rangle}}$ km/sec; $(l_0,b_0)=( 292\pm21, 27 \pm 15)^\circ$} \\
\multicolumn{14}{l}{\vspace{0.1in} $^{(b)}$ $d\sim 370-540
h_{70}^{-1}$Mpc; $(V_x,V_y,V_z)=
    (428 \pm 375 , -1,029 \pm 323, 575 \pm 316 )\times\frac{+0.3\mu
      K}{\sqrt{\langle{C}_{1,100}\rangle}}$ km/sec; $V_{\rm Bulk}=(1,254\pm
    328 )\times\frac{+0.3\mu K}{\sqrt{\langle{C}_{1,100}\rangle}}$ km/sec;
    $(l_0,b_0)=(293\pm 20, 27\pm 15)^\circ$} \\
\hline
0.20 & 0.5--1 & 238 & 0.093/0.089 & $-2.5\pm2.1, -1.3 \pm 1.8, 1.0 \pm 1.8$ & $3.0\pm2.0$ & 3.5 & 0.2828 & 0.2390 & --2.9 & --2.2  & --1.1 & --0.3 & --0.2 \\
0.20 & 1--2 & 248 & 0.122/0.123 & $0.1\pm2.0, -1.8 \pm 1.8, -0.3\pm 1.7$ & $1.8\pm1.8$ & 4.4 & 0.3231 & 0.2835 & --5.1 & --1.8  & --0.3 & 0.5 & 1.0 \\
0.20 & $>2$ & 208 & 0.140/0.151 & $3.6\pm2.2, -5.8 \pm1.9, 4.5\pm1.9$ & $8.1\pm2.0$ & 6.6 & 0.4644 & 0.4218 & --9.3 & --5.5  & --1.9 & 0.4 & 1.1 \\
\hline \\
\multicolumn{14}{l}{$^{(a)}$ $d\sim 380-650 h_{70}^{-1}$Mpc;
$(V_x,V_y,V_z)=
  (213 \pm 341, -872\pm 294, 529 \pm 287)\times\frac{+0.3\mu
    K}{\sqrt{\langle{C}_{1,100}\rangle}}$ km/sec; $V_{\rm Bulk}=(1,042 \pm
  295)\times\frac{0.3\mu K}{\sqrt{\langle{C}_{1,100}\rangle}}$ km/sec;
  $(l_0,b_0)=(284\pm 24, 30 \pm 16)^\circ$} \\
\multicolumn{14}{l}{\vspace{0.1in} $^{(b)}$ $d\sim 380-650
h_{70}^{-1}$Mpc;
  $(V_x,V_y,V_z)= (248 \pm 337, -880\pm 291, 538 \pm 284)\times\frac{0.3\mu
    K}{\sqrt{\langle{C}_{1,100}\rangle}}$ km/sec; $V_{\rm Bulk}=(1,061 \pm
  292)\times\frac{0.3\mu K}{\sqrt{\langle{C}_{1,100}\rangle}}$ km/sec;
  $(l_0,b_0)=(286\pm 23, 30 \pm 15)^\circ$} \\
\hline
0.25 & 0.5--1 & 240 & 0.094/0.090 & $-2.3\pm2.1, -1.1 \pm 1.8, 0.9 \pm 1.8$ & $2.7\pm2.0$ & 3.5 & 0.2848 & 0.2444 & --2.8 & --2.1  & --1.0 & --0.3 & --0.1 \\
0.25 & 1--2 & 276 & 0.133/0.133 & $-0.2 \pm 2.0, -1.4\pm 1.7, 0.7 \pm 1.6$ & $1.6\pm1.7$ & 4.4 & 0.3162 & 0.2806 & --5.8 & --2.3  & --0.8 & -0.1 & 0.3 \\
0.25 & $>2$ & 322 & 0.169/0.176 & $3.7\pm1.8,-4.1\pm1.5,4.1\pm1.5$ &  $6.9\pm1.6$  & 6.6 & 0.4434 & 0.4160 & --6.9 & --4.6  & --2.3 & --0.6 & 0.2 \\
\hline \\
\multicolumn{14}{l}{$^{(a)}$ $d\sim 385-755 h_{70}^{-1}$Mpc;
$(V_x,V_y,V_z)=
  (313\pm 308, -707\pm 265, 643\pm 259)\times\frac{+0.3\mu
    K}{\sqrt{{\langle{C}}_{1,100}\rangle}}$ km/sec; $V_{\rm
    Bulk}=(1,005\pm267)\times\frac{0.3\mu
    K}{\sqrt{\langle{C}_{1,100}\rangle}}$ km/sec; $(l_0,b_0)=(296 \pm 29, 39
  \pm 15)^\circ$} \\
\multicolumn{14}{l}{\vspace{0.1in} $^{(b)}$ $d\sim 385-755
h_{70}^{-1}$Mpc;
  $(V_x,V_y,V_z)= (352\pm 304, -713\pm 262, 652\pm 256)\times\frac{+0.3\mu
    K}{\sqrt{{\langle{C}}_{1,100}\rangle}}$ km/sec; $V_{\rm
    Bulk}=(1,028\pm265)\times\frac{0.3\mu
    K}{\sqrt{\langle{C}_{1,100}\rangle}}$ km/sec; $(l_0,b_0)=(296 \pm 28, 39
  \pm 14)^\circ$} \\
\hline
\enddata
\tablecomments{ }
\end{deluxetable}
 \clearpage
\begin{figure}
\plotone{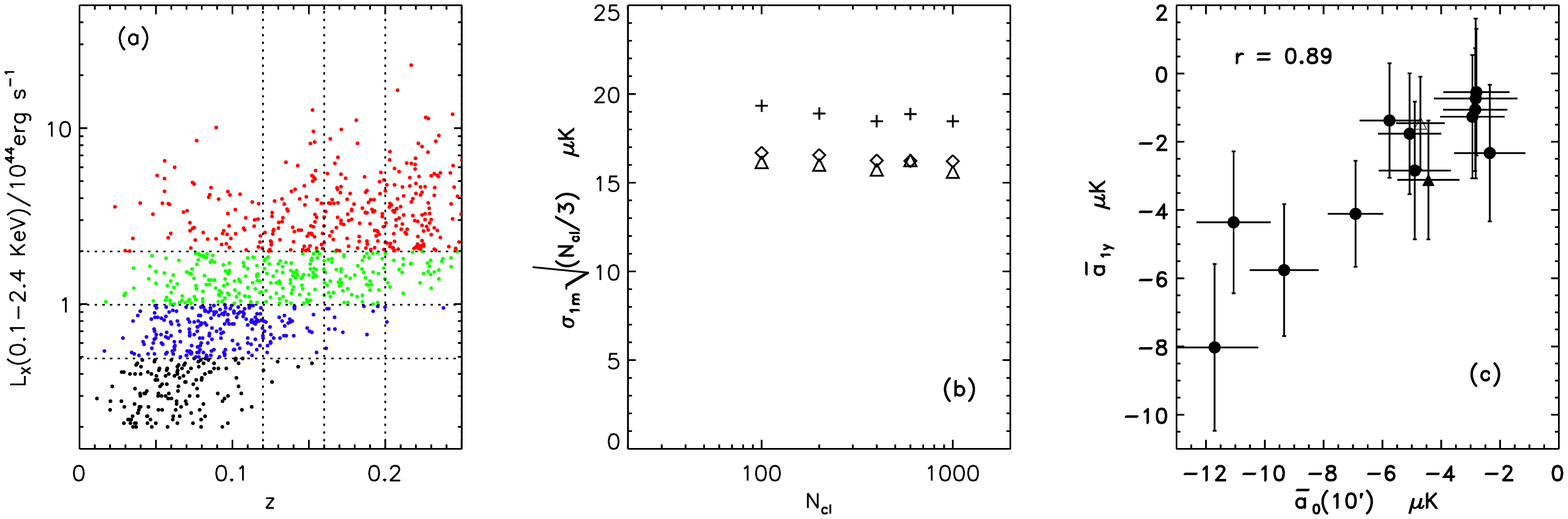} \caption{ \small{ }} \label{fig:f1}
\end{figure}

\clearpage

\begin{figure}
\plotone{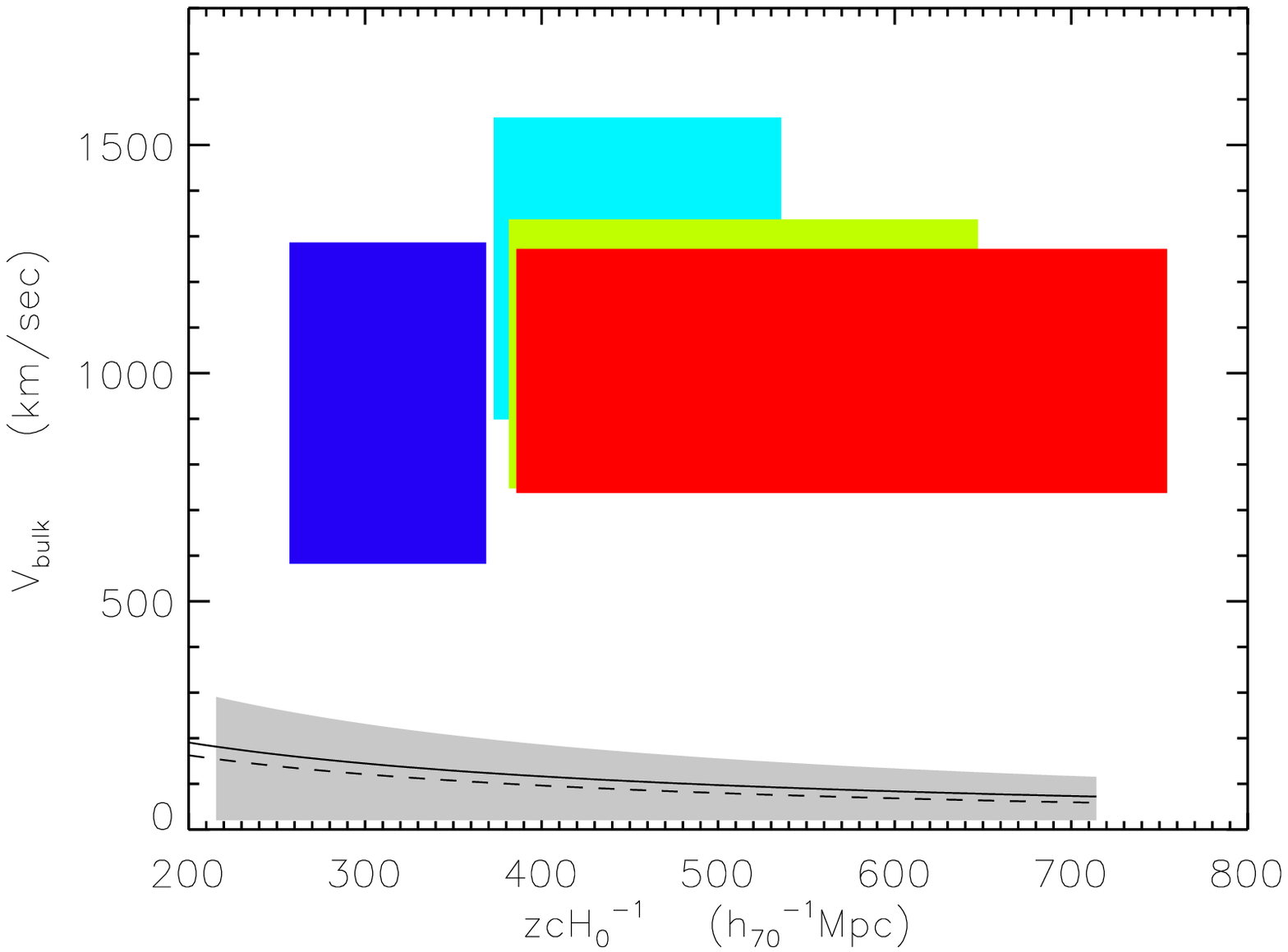} \caption{ \small{ }} \label{fig:f2}
\end{figure}

\clearpage

\section*{Figure and table captions:}

{\bf Table 1}: : Columns are: (1) - the limit of the cumulative
$z$-bin. (2) luminosity range of the differential $L_X$-bin. (3)
the number of clusters per bin. (4) mean/median redshift in the
bin. (5) Dipole coefficients, averaged over the eight WMAP DA's
over the clusters in the bin with 1-$\sigma$ errors,
$\sigma_{1m}$. (6) Dipole amplitude, $\sqrt{C_1} = \sqrt{\sum_m
a_{1m}^2}$. The error on the dipole amplitude is derived as
$\sigma_1^2=\sum_m (\partial \sqrt{C_1}/\partial a_{1m})^2
\sigma_{1m}^2=\sum_m a_{1m}^2\sigma_{1m}^2/C_1$. The binning in
luminosity was designed, wherever possible, to bin different
z-bins by the same luminosity range. Note that $a_{1y}$ are always
negative. When we select 598 clusters out to $z\leq 0.25$ with
$L_X\geq 10^{44}$ erg/sec the dipoles are consistent with the
brightest $L_X$-bin: $(\bar{a}_{1x},
\bar{a}_{1y},\bar{a}_{1z})=(-0.3\pm 1.2, -2.5 \pm 1.1, 2.4 \pm
1.1)\mu$K. (7) Central optical depths, $\tau_0\equiv \sqrt{\pi}
\sigma_{T} n_{e,0}R_{\rm core}$, averaged over clusters in the bin
derived from our cluster catalog as described in the text. (8)
Calibration factors, $\sqrt{C_{1,100}}$, for clusters in the given
bin at 5$^\prime$ radial distance from the cluster centers and at
1 X-ray extent for $\beta$-model with $\beta=2/3$. (9) Measured
monopole,
over the fixed aperture with the radius shown, after averaging over all DA's.\\
The row at the end of each $z$-bin sums up the bulk flow
parameters (scale, components, amplitude and direction to the axis
of motion) assuming a coherent motion for all the $L_X$-bins. The
depth is defined as $d\equiv z_{\rm median} cH_0^{-1}$. \\
$^{(*)}$ There are only five clusters in this $L_X$-range at
$z>0.12$, so for brevity this configuration's parameters are not
repeated the remaining $z$-bins, although for completeness it is
included in bulk flow evaluations $^{(b)}$.\\
$^{(a)}$ The bulk flow is derived using column 8 parameters
without the lowest $L_X$-clusters which do not extend beyond $z=0.12$ $^{(*)}$ (see text).\\
$^{(b)}$ The bulk flow is derived using column 7 parameters
including the lowest $L_X$-clusters $^{(*)}$ (see text).\\

{\bf Figure 1}: (a) $z$-distribution of clusters in the $L_X$-bins
used in the analysis. Black/blue/green/red colors correspond to
$L_X = (0.2-0.5, 0.5-1, 1-2, >2) \times 10^{44}$ erg/sec
respectively. (b) Standard deviations from simulations of random
pseudo-clusters vs $N_{\rm cl}$. Pluses/diamonds/triangles
correspond to $a_{1x}/a_{1y}/a_{1z}$. (c) Correlation between the
$y$-component of the dipole and the central monopole from Table 1
(filled circles). Both quantities increase in amplitude with $L_X$
because the optical depth and central temperature increase for
more massive clusters: KSZ dipole scales as $\tau$ and the
monopole as $\tau T_X$. The linear correlation coefficient for the
circles is shown in the upper left. Triangles correspond to
differential $z$-configuration, $0.12 < z \leq 0.25$, where we
still have enough luminous clusters for reasonable S/N.
Open/filled triangles correspond to $L_X> (1,2)\times 10^{44}$
erg/sec with 418/260 clusters. When the triangles are included the
linear correlation coefficient becomes 0.85.

{\bf Figure 2}: Bulk velocity vs depth from Table 1:
blue/cyan/grean/red corresponds to $z\leq 0.12/0.16/0.2/0.25$;
parameters for fits (a) are chosen for brevity. Solid/dashed lines
correspond to the rms bulk velocity for the concordance
$\Lambda$CDM model for top-hat/Gaussian windows. Black-shaded
regions shows the 95\% confidence level of the model (see KABKE1
for details).

\end{document}